\documentclass[preprint,showpacs,preprintnumbers,amsmath,amssymb,footnotepage]{revtex4-1}
\usepackage{subfigure}
.tfm scaled 1000
\usepackage{amsmath,amscd,amssymb,euscript,eufrak}
\usepackage{graphicx}
\usepackage{graphics}
\usepackage{epsfig}
\usepackage{ifthen}
\usepackage{verbatim}
\usepackage{color}

\usepackage{makeidx}   
\DeclareMathAlphabet{\mathpzc}{OT1}{pzc}{m}{it}
\def\beq{\begin{equation}}
\def\eeq{\end{equation}}

\newtheorem{theo}{{\em Theorem}}
\newtheorem{defin}{{\em Definition}}
\begin{document}
\def\ss #1{ \hskip -1.5pt #1 \hskip -1.5pt }
\def\pl{\partial}
\def\ssquare{{\scriptscriptstyle \square}}
\def\ssq{{\scriptscriptstyle \square}}

\newcommand{\be}{\begin{equation}}
\newcommand{\ee}{\end{equation}}
\newcommand{\ben}{\begin{eqnarray}}
\newcommand{\een}{\end{eqnarray}}
\newcommand{\pp}{\prime}
\newcommand{\nn}{\nonumber}

\newcommand{\ov}{\overline}
\newcommand{\kn}{| n \rangle}
\newcommand{\bn}{ \langle n |}
\newcommand{\til}{\tilde}
\newcommand{\iii}{\'{\i}}

\newcommand{\kc}{| c \rangle}
\newcommand{\bc}{ \langle c |}
\newcommand{\kz}{| z \rangle}
\newcommand{\bz}{ \langle z|}
\newcommand{\kcz}{| \til{c}_z \rangle}
\newcommand{\bcz}{ \langle \til{c}_z |}

\newcommand{\lan}{  \lambda}
\newcommand{\kp}{| \pp \rangle}
\newcommand{\bp}{ \langle  \p2  |}
\newcommand{\ATA}{\hat{A}^\dagger \hat{A}}
\newcommand{\AAT}{  \hat{A} \hat{A}^\dagger}
\newcommand{\AT}{\hat{A}^\dagger}
\newcommand{\AO}{\hat{A}}
\newcommand{\sn}{\sum^M_{n=1}}
\newcommand{\snn}{\sum^N_{n=1}}
\newcommand{\RN}{{\cal {R}}^N}
\newcommand{\RM}{{\cal {R}}^M}
\newcommand{\kpn}{ | \psi_n \rangle}
\newcommand{\bpn}{\langle \psi_n|}
\newcommand{\kfn}{ | \phi_n \rangle}
\newcommand{\bfn}{\langle \phi_n|}
\newcommand{\kfo}{ | \ov{\phi}_n \rangle}
\newcommand{\bfo}{\langle \ov{ \phi}_n|}

\newcommand{\kf}{ | \ov{f} \rangle}
\newcommand{\kfe}{ | f^o \rangle}

\newcommand{\ton}{ \,;\, n=1,\ldots,N}
\newcommand{\toi}{ \,;\, i=1,\ldots,M}

\newcommand{\NC}{\hat{N}^\bot}
\newcommand{\NU}{Null(\AO)}
\newcommand{\NUC}{Null^\bot(\AO)}
\newcommand{\RA}{ Range(\AO)}
\newcommand{\PN}{\hat{P}_{N}}
\newcommand{\PNC}{\hat{P}_{N^{\bot}}}
\newcommand{\PR}{\hat{P}_{R}}
\def\stackunder#1#2{\mathrel{\mathop{#2}\limits_{#1}}}
\font\Bbb =msbm10  scaled \magstephalf
\def\id{{\hbox{\Bbb I}}}
\renewcommand{\theenumi}{\alph{enumi}}
\newcommand{\ia}{\'{\i}}
\newcommand{\nd}{\noindent}

\def\fa{\mathfrak{a}}
\def\fc{\mathfrak{c}}
\def\fd{\mathfrak{d}}
\def\fb{\mathfrak{b}}
\def\fj{\mathfrak{j}}
\def\fg{\mathfrak{g}}
\def\fh{h}
\def\fk{\mathfrak{k}}
\def\fr{\mathfrak{r}}
\def\fs{\mathfrak{s}}
\def\fgl{\mathfrak{gl}}
\def\fsl{\mathfrak{sl}}
\def\fsu{\mathfrak{su}}
\def\fsp{\mathfrak{sp}}
\def\fso{\mathfrak{so}}
\def\fu{u}

\title{A geometric approach to the distribution of quantum states in bipartite physical systems}
\author{J Batle$^{1}$}

\email{E-mail address: jbv276@uib.es}
\affiliation{ $^1$Departament de F\'{\i}sica, Universitat de les Illes Balears,
 07122 Palma de Mallorca, Balearic Islands, Europe  \\\\}
 
\date{\today}

\begin{abstract}

Any set of pure states living in an given Hilbert space possesses a natural and unique metric --the Haar measure-- on the group $U(N)$ of unitary matrices. However, 
there is no specific measure induced on the set of eigenvalues $\Delta$ of any density matrix $\rho$. Therefore, a general approach to the global properties 
of mixed states depends on the specific metric defined on $\Delta$. 
In the present work we shall employ a simple measure on $\Delta$ that has the advantage of possessing 
a clear geometric visualization whenever discussing how arbitrary states are distributed according to some measure of mixedness. The degree of mixture will be that of the participation ratio $R=1/Tr(\rho^2)$ and the concomitant maximum eigenvalue $\lambda_m$. The cases studied will be the qubit-qubit system and the qubit-qutrit system, whereas some discussion will be made on higher-dimensional bipartite cases in both the $R$-domain and the $\lambda_m$-domain.

\end{abstract}
\date{\today}
\maketitle

%
%

\section{Introduction}

 The amount of entanglement and the purity of quantum states of composite
  systems exhibit a dualistic relationship. As the degree of
  mixture increases, quantum states tend to have a smaller
  amount of entanglement. In the case of two-qubits systems,
  states with a large enough degree of mixture are always
  separable \cite{ZHS98}. A detailed knowledge of the relation between the degree
  of mixture and the amount of entanglement is essential in order to
  understand the limitations that mixture imposes on quantum information
  processes such as quantum teleportation or quantum computing.
  To study the relationship between entanglement and mixture
  we need quantitative measures for these two quantities.
  The entanglement of formation provides a natural quantitative
  measure of entanglement with a clear physical motivation.
  As for mixedness, there are several measures of mixture that can
  be useful within the present context. The von Neumann measure

  \be \label{slog}
  S \, = \, - Tr \left( \hat \rho \ln \hat \rho \right),
  \ee

  \noindent is important because of its relationship with the thermodynamic
  entropy. On the other hand, the so called participation ratio,

  \be \label{partrad}
  R(\hat \rho) \, = \, \frac{1}{Tr(\hat \rho^2)},
  \ee

  \noindent
  is particularly convenient for calculations \cite{ZHS98,MJWK01}. Another measure for 
  mixedness can be found in the maximum eigenvalue of a density matrix $\lambda_m$, 
  which is in turn a monotonically increasing function for the Renyi entropy $SR_{\infty}$.
  
  Given a particular way to explore the space of both pure and mixed states in bipartite systems, it is possible to 
  provide a clear physical geometric insight into the problem of how states distribute according to their degree of mixture, which is the main 
  subject of the present work.
  
  This paper is organized as follows: in Section II we introduce the nature of the measures defined on the set of eigenvalues 
  and unitary matrices. Section III analyzes 2x2 systems and how their concomitant states are distributed according to 
  $R$ and $\lambda_m$. Section IV provides a further analytic approach for 2x3 and additional bipartite systems. In Section V 
  we discuss the implications of the non-uniqueness of a general measure for mixed states and the corresponding 
  geometric implications into our problem. Finally, some conclusions are drawn in Section VI.

\section{Measures on the set of eigenvalues and unitary matrices}

In order to perform a survey of the properties of arbitrary (pure and mixed) states of 
the concomitant state-space ${\cal
S}$, it is necessary to introduce an appropriate measure $\mu $
on this space. Such a measure is needed to compute volumes within ${\cal S}$,
as well as to determine what is to be understood by a uniform distribution of
states on ${\cal S}$.  The natural measure that we are going to adopt here was first considered in 
Refs. \cite{ZHS98,Z99}.
An arbitrary (pure or mixed) state $\rho$ of a quantum system
described by an $N$-dimensional Hilbert space can always be
expressed as the product of three matrices,

\be \label{udot} \rho \, = \, U D[\{\lambda_i\}] U^{\dagger}. \ee

\noindent Here $U$ is an $N\times N$ unitary matrix and
$D[\{\lambda_i\}]$ is an $N\times N$ diagonal matrix whose
diagonal elements are $\{\lambda_1, \ldots, \lambda_N \}$, with $0
\le \lambda_i \le 1$, and $\sum_i \lambda_i = 1$.
The group of unitary matrices $U(N)$ is
endowed with a unique, uniform measure: the Haar measure $\nu$
\cite{PZK98}. On the other hand, the $N$-simplex $\Delta$,
consisting of all the real $N$-uples $\{\lambda_1, \ldots,
\lambda_N \}$ appearing in (\ref{udot}), is a subset of a
$(N-1)$-dimensional hyperplane of ${\cal R}^N$. Consequently, the
standard normalized Lebesgue measure ${\cal L}_{N-1}$ on ${\cal
R}^{N-1}$ provides a natural measure for $\Delta$. The
aforementioned measures on $U(N)$ and $\Delta$ lead then to a
natural measure $\mu $ on the set ${\cal S}$ of all the states of
our quantum system \cite{ZHS98,Z99,PZK98}, namely,

\be \label{memu1}
 \mu = \nu \times {\cal L}_{N-1}.
 \ee

 \noindent
  All our present considerations are based on the assumption
 that the uniform distribution of states of a quantum system
 is the one determined by the measure (\ref{memu1}). Thus, in our
 numerical computations, we are going to randomly generate
 states according to the measure (\ref{memu1}). The quantities $\mu_i$ computed with a Monte 
 Carlo procedure have an associated error which is on the type 
 $t_{M-1,\alpha/2}\frac{\sigma_x}{\sqrt{M-1}}$, where $M$ is the number 
 of generated states, $t_{M-1,\alpha/2}$ is the value corresponding to 
 the Student distribution with $M-1$ degrees of freedom, computed with a 
 certain desired accuracy $1-\alpha$, and $\sigma_x$ is the usual computed 
 standard deviation. Therefore, if we seek a result with an error say less than 
 $10^{-3}$ units, we have to generate a number of points $M$ around 
 10 or 100 million. If not stated explicitly, from now on all quantities computed 
 are exact up to the last digit.  
 
 The applications that have appeared so far in quantum information theory, 
in the form of dense coding, teleportation, quantum cryptography and specially 
in algorithms for quantum computing (quantum error correction codes for instance), 
deal with finite numbers of qubits. A quantum gate which acts upon these qubits 
or even the evolution of 
that system is represented by a unitary matrix $U(N)$, with $N=2^n$ being the 
dimension of the associated Hilbert space ${\cal H}_N$. The state $\rho$ describing 
a system of $n$ qubits is given by a hermitian, positive-semidefinite ($N \times N$) 
matrix, with unit trace. In view of these facts, it is natural 
to think that an interest has appeared in the {\it quantification} of 
certain properties of these systems, most of the 
times in the form of the characterization of a certain state $\rho$, 
described by $N \times N$ matrices of finite size. Natural applications arise 
when one tries to simulate certain processes through random matrices, 
whose probability distribution ought to be described accordingly.

\subsection{The Haar measure}

As stated before, in the space of pure states, with $|\Psi\rangle \in {\cal H}_N$, 
there is a natural candidate measure, induced by 
the {\bf Haar measure} on the group ${\cal U}(N)$ of unitary matrices. 
In mathematical analysis, the Haar measure \cite{Haar33} 
is known to assign an ``invariant 
volume" to what is known as subsets of locally compact topological groups. 
In origin, the main objective was to construct a measure invariant under the 
action of a topological group \cite{Mehta90}. Here we present the formal 
definition \cite{Conway90}: given a locally compact topological group 
$G$ (multiplication is the group operation), consider a $\sigma$-algebra $Y$ 
generated by all compact subsets of $G$. 
If $a$ is an element of $G$ and $S$ is a set in $Y$, then the set 
$aS = $ $\{$ $as : s \in S$ $\}$ also belongs to $Y$. A measure $\mu$ on $Y$ will be 
letf-invariant if $\mu(aS)=\mu(S)$ for all $a$ and $S$. Such an invariant measure 
is the Haar measure $\mu$ on $G$ (it happens to be both left and 
right invariant). In other words \cite{Haarsimetria}, the Haar measure defines 
the unique invariant integration measure for Lie groups. It implies that a 
volume element d$\mu(g)$ is identified by defining the integral of a function 
$f$ over $G$ as $\int_G f(g) d\mu(g)$, being left and right invariant 

\begin{equation}
\int_G f(g^{-1}x) d\mu(x)\,=\,\int_G f(x g^{-1}) d\mu(x)\,=\,\int_G f(x) d\mu(x).
\end{equation}

\noindent The invariance of the integral follows from the concomitant invariance 
of the volume element d$\mu(g)$. It is plain, then, that once d$\mu(g)$ is fixed 
at a given point, say the unit element $g=e$, we can move performing a 
left or right translation. Suppose that the map $x \rightarrow g(x)$ defines the 
action of a left translation. We have $x^i \rightarrow y^i(x^j)$, with $x^i$ being 
the coordinates in the vicinity of $e$. Assume, also, that d$x^1 ...$d$x^n$ defines 
the volume element spanned by the differentials d$x^1$, d$x^2$, ..., d$x^n$ at point 
$e$. It follows then that the volume element at point $g$ is given by 
d$\mu(g)=|J|^{-1}$d$x^1 ...$d$x^n$, where $J$ is the Jacobian of the previous 
map evaluated at the unit element $e$: $J=\frac{\delta(y^1...y^n)}{\delta(x^1...x^n)}$. 
In a right or left translation, both d$x^1 ...$d$x^n$ and $|J|$ are multiplied by the 
same Jacobian determinant, preserving invariance of d$\mu(g)$. The Lie groups 
also allow an invariant metric and d$\mu(g)$ is just the volume element 
$\sqrt{g}$d$x^1 ...$d$x^n$.

We do not gain much physical insight with these definitions of the Haar measure and its 
invariance unless we identify 
$G$ with the group of unitary matrices ${\cal U}(N)$, the element $a$ with a 
unitary matrix $U$ and $S$ with subsets of the group of unitary matrices ${\cal U}(N)$, 
so that given a reference state $|\Psi_0\rangle$ and a unitary matrix $U \in {\cal U}(N)$, 
we can associate a state $|\Psi\rangle_0=U|\Psi_0\rangle$ to $|\Psi_0\rangle$.
Physically what is required is a probability measure $\mu$ invariant under 
unitary changes of basis in the space of pure states, that is, 

\begin{equation}
P^{(N)}_{Haar}(U\,|\Psi\rangle)\,=\,P^{(N)}_{Haar}(|\Psi\rangle).
\end{equation}

\noindent These requirements can only be met by the Haar measure, which is 
rotationally invariant.

\section{Analytical approach for 2x2 systems. Generation of states}

The two-qubits case ($N=2 \times 2$) is the simplest quantum mechanical 
system that exhibits the feature of quantum entanglement. The relationship 
between entanglement and mixedness has been described intensively in the literature. 
One given aspect is that as we increase the degree               
of mixture, as measured by the so called participation ratio 
$R=1/$Tr[$\rho^2$], the entanglement diminishes (on average). 
As a matter of fact, if the state is mixed enough, that state will have 
no entanglement at all. This is fully consistent with the fact 
that there exists a special class of mixed states which have maximum 
entanglement for a given $R$ \cite{MJWK01} 
(the maximum entangled mixed states MEMS). 
These states have been recently reported to be achieved in the 
laboratory \cite{MEMSexp} using pairs of entangled photons. 
Thus for practical or purely theoretical purposes, it may happen 
to be relevant to generate mixed states of two-qubits with 
a given participation ratio $R$. It may represent an excellent tool 
in the simulation of algorithms in a given quantum circuit: 
as the input pure states go through the quantum gates, they interact with the 
environment, so that they become mixed with some $R$. This 
degree of mixture $R$, which varies with the number of iterations, can be used 
as a probe for the evolution of the degradation of the entanglement present 
between any two qubits in the circuit. Different evolutions of the degree of mixture on 
the output would shed some light on the optimal architecture of the circuit 
that has to perform a given algorithm.

Here we describe a numerical recipe to randomly generate two-qubit states, according 
to a definite measure, and with a given, fixed value of $R$. Suppose 
that the states $\rho$ are generated according to the product measure 
$\mu = \nu \times {\cal L}_{N-1}$ (\ref{memu1}), where $\nu$ is the Haar measure 
on the group of unitary matrices ${\cal U}(N)$ and the Lebesgue measure 
${\cal L}_{N-1}$ on ${\cal R}^{N-1}$ provides a reasonable measure for
the simplex of eigenvalues of $\rho$. In this case, the numerical procedure 
we are about to explain owes its efficiency 
to the following {\it geometrical picture} which is {\it valid only 
if the states are supposed to be distributed according to 
measure} (\ref{memu1}). We shall identify the simplex $\Delta $ with a regular 
tetrahedron of side length 1, in ${\cal R}^3$, centred at the origin. Let
  ${\bf r}_i$ stand for the vector positions of the tetrahedron's
  vertices. The tetrahedron is oriented in such a way that the vector
  ${\bf r}_4$ points towards the positive $z$-axis and the vector
  ${\bf r_2}$ is contained in the $(x,z)$-semiplane corresponding to
  positive $x$-values. The positions of the tetrahedron's vertices correspond to 
  the vectors

\begin{eqnarray} 
\bf{r_1} &=& (-\frac{1}{2\sqrt{3}},-\frac{1}{2},-\frac{1}{4}\sqrt{\frac{2}{3}}) \nonumber \\
\bf{r_2} &=& (\frac{1}{\sqrt{3}},0,-\frac{1}{4}\sqrt{\frac{2}{3}}) \nonumber \\
\bf{r_3} &=& (-\frac{1}{2\sqrt{3}},\frac{1}{2},-\frac{1}{4}\sqrt{\frac{2}{3}}) \nonumber \\
\bf{r_4} &=& (0,0,\frac{3}{4}\sqrt{\frac{2}{3}}).
\end{eqnarray}

\noindent The mapping connecting the points
  of the simplex $\Delta $ (with coordinates $(\lambda_1,\ldots, \lambda_4)$)
  with the points $\bf r$ within tetrahedron is given by the equations

  \begin{eqnarray} \label{tetra1}
  \lambda_i \, &=& \, 2({\bf r}\cdot {\bf r}_i ) \, + \, \frac{1}{4}
  \,\,\,\, i=1, \dots, 4, \cr
  {\bf r} \, &=& \, \sum_{i=1}^4 \lambda_i {\bf r}_i
  \end{eqnarray}

  \noindent The degree of mixture is characterized by the
  quantity $R^{-1} \equiv Tr(\rho^2) = \sum_i \lambda_i^2$. This
  quantity is related to the distance $r=\mid {\bf r} \mid$
  to the centre of the tetrahedron $T_{\Delta}$ by

  \begin{equation} \label{tetra3}
  r^2 \, = \, -\frac{1}{8} \, + \, \frac{1}{2} \sum_{i=1}^4 \lambda_i^2.
  \end{equation}

 \noindent Thus, the states with a given degree of mixture lie on the
 surface of a sphere $\Sigma_r$ of radius $r$ concentric with the
 tetrahedron $T_{\Delta}$. To choose a given $R$ is tantamount to define 
a given radious of the sphere. There exist three different possible regions 
(see Fig. 1):

\begin{figure}[ht]
\centering   \includegraphics[scale= 1]{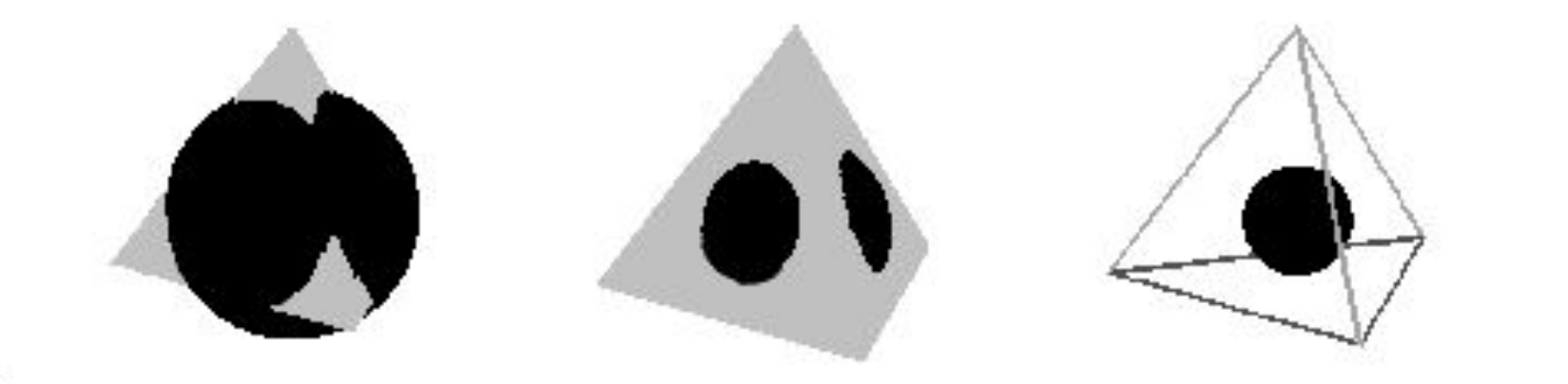}
\caption{Geometric evolution of a growing sphere inside a tetrahedron, depicting the distribution nature of states $\rho$ generated 
according to the measure (\ref{nu}). See text for details.}
\end{figure}

\begin{itemize}

\item region I: $r \in [0, h_1]$ ($R \in [4,3]$), where
$h_1 \equiv h_c={1 \over 4 }\sqrt{2 \over {3}}$ is the radius of a sphere
tangent to the faces of the tetrahedron $T_{\Delta}$. In this case
the sphere $\Sigma_r$ lies completely within the tetrahedron
$T_{\Delta}$. Therefore we only need to generate at random points over its 
surface. The cartesian coordinates for the sphere are given by

\begin{eqnarray} \label{spher}
x_1 &=& r \, \sin\theta\, \cos\phi \nonumber \\
x_2 &=& r \, \sin\theta\, \sin\phi \nonumber \\ 
x_3 &=& r \, \cos\theta, 
\end{eqnarray}

\noindent Denoting {\sf rand\_{}u()} a random number uniformly distributed 
between 0 an 1, the random numbers $\phi=2\pi${\sf rand\_{}u()} and 
$\theta=\arccos(2${\sf rand\_{}u()}$-1)$ (its probability distribution being 
$P(\theta)=\frac{1}{2}\sin(\theta)$) define an arbitrary state $\rho$ on the 
surface {\it inside} $T_{\Delta}$. The angle $\theta$ is defined between the 
centre of the tetrahedron (the origin) and the vector ${\bf r_4}$, and any point 
aligned with the origin. Substitution of ${\bf r}=(x_1,x_2,x_3)$ 
in (\ref{tetra1}) provides us with 
the eigenvalues $\{$$\lambda_i$$\}$ of $\rho$, with the desired $R$ as 
prescribed 
by the relationship (\ref{tetra3}). With the subsequent application of the 
unitary matrices $U$ we obtain a random state $\rho = U D(\Delta) U^{\dag}$ 
distributed according to the usual measure $\mu = \nu \times {\cal L}_{N-1}$.

\item region II: $r \in [h_1, h_2]$ ($R \in [3,2]$), where 
$h_2 \equiv \sqrt{h^{2}_{c}+(\frac{D}{2})^2}={\sqrt{2}\over 4}$ denotes
 the radius of a sphere which is tangent to the sides of the tetrahedron
 $T_{\Delta}$. Contrary to the previous case, part of the surface of the 
sphere lies outside the tetrahedron. This fact means that we are able to 
still generate the states $\rho$ as before, provided we reject those ones with 
negative weights $\lambda_i$.
									      
\item region III: $r \in [h_2, h_3]$ ($R \in [2,1]$), where
  $h_3 \equiv \sqrt{h^{2}_{c}+D^2}={\sqrt{6}\over 4}$ is the radius of 
a sphere passing through the vertices of $T_{\Delta}$. The generation 
of states is a bit more involved in this case. Again 
$\phi=2\pi${\sf rand\_{}u()}, but the available 
angles $\theta$ now range from $\theta_c(r)$ to $\pi$. It can be shown that 
$w\equiv\cos(\theta_c)$ results from solving the equation 
$3r^2 w^2 - \sqrt{\frac{3}{2}}r w + \frac{3}{8}-2r^2 = 0$. Thus, 
$\theta(r)=\arccos(w(r))$, with $w(r)=\cos\theta_c(r) + 
(1-\cos\theta_c(r))${\sf rand\_{}u()}. 
Some states may be unacceptable ($\lambda_i<0$) still, but the vast majority 
are accepted.

\end{itemize}

Combining these three previous regions, we are able to generate arbitrary 
mixed states $\rho$ endowed with a given participation ratio $R$.

\subsection{R-domain}

  In this case the degree of mixture is characterized by the
  quantity $\omega_2 = Tr(\hat \rho^2)=\sum_i p_i^2$. 
  This quantity is related to the distance $r=\mid {\bf r} \mid$
  to the centre of the tetrahedron $T_{\Delta}$ by

  \be \label{tetra2}
  r^2 \, = \, -\frac{1}{8} \, + \, \frac{1}{2} \omega_2.
  \ee

 \noindent
 Thus, the states with a given degree of mixture lie on the
 surface of a sphere of radius $r$ concentric with the
 tetrahedron $T_{\Delta}$.


The volume associated with states endowed with a value of
$\omega_2$ lying within a small interval $dw_2$ is clearly
associated with the volume $dV$ of the subset of points in
$T_{\Delta}$ whose distances to the centre of $T_{\Delta}$ are
between $r$ and $r+dr$, with $r dr = \omega_2 \, d\omega_2$. Let
$\Sigma_r$ denote the sphere of radius $r$ concentric with
$T_{\Delta}$. The volume $dV$ is then proportional to the area
$A(r)$ of the part of $\Sigma_r$ which lies within $T_{\Delta}$.
In order to compute the aforementioned area, it is convenient to
separately consider three different ranges for the radius $r$.

Let us first consider the range of values  $r \in [0, h_1]$, where
$h_1={1 \over 4 }\sqrt{2 \over {3}}$ is the radius of a sphere
tangent to the faces of the tetrahedron $T_{\Delta}$. In this case
the sphere $\Sigma_r$ lies completely within the tetrahedron
$T_{\Delta}$. Thus, the area we are interested in is just the area
of the sphere,

\begin{equation} \label{a1}
A_{I}(r) = 4 \pi r^{2}. 
\end{equation}

We now consider a second range of values of the radius, $r \in
[h_1, h_2]$, where  $h_2 ={\sqrt{2}\over 4}$ denotes
 the radius of a sphere which is tangent to the sides of the tetrahedron
 $T_{\Delta}$. In this case, the area of the portion of $\Sigma_r$
 which lies within $T_{\Delta}$ is

\begin{equation} \label{a2}
A_{II}(r) = 4 \pi \left[r^{2}-2r(r-h_1)\right]. 
\end{equation}

 Finally, we consider the range of values  $r \in [h_2, h_3]$, where
  $h_3={\sqrt{6}\over 4}$ is the radius of a sphere passing through the vertices
 of $T_{\Delta}$. This cas is, by far, the most intricate one, where many methods borrowed 
 from spherical trigonometry are employed. 
 In this case the area $A_{III}$ of the part of the sphere
 $\Sigma_r$ lying within $T_{\Delta}$ is

\begin{equation} \label{a3}
A_{III}(r) = 4 (S_{A}-3S_{B}), 
\end{equation}

\noindent where

\ben S_A &=& r^2 (3 \beta - \pi), \cr
 S_B &=& r^2 \left[ h (-\pi + 2 \sin^{-1} (C_1 C_2)) +
 2 \sin^{-1} \sqrt{1 - C^2_1 C^2_2 \over 1
+ C^2_2} \right]. \een

\noindent The quantities appearing in the right hand sides of the
above expressions are defined by

\be
 \beta = \cos^{-1} \left[{\cos A - \cos^2 A \over \sin^2
A}\right]; \,\,\,\, A = 2 \sin^{-1} (D_1/r);  \,\,\,\, D_1 = {1
\over 2} \left({1\over 2} - \sqrt{r^2 - {1 \over 8}}\right),
 \ee

 \noindent
 and

\ben
h &=& h_1/r; \ \  C_1 ={h \over \sqrt{1 - h^2}}; \ \ C_2 = {C_B
\over \sqrt{1 - C^2_B} }; \cr
C_B &=& \sqrt{D^2_2 - D^2_1 \over r^2 - D^2_1}; \ \ D_2 = r \sqrt{1 - h^2}. 
\een

\noindent
 Using the relation between $r$ and the participation rate
$R = 1/ Tr(\rho^2)$,

\be \label{radiopart}
 r^2 = -  {1 \over 8} + {1 \over 2 R},
\ee

\noindent
 we analytically obtained the probability $F(R)$ of
finding a quantum state with a participation rate $R$,

\be
\label{ffrr} F(R) \, = \, f(r)   \left| \frac{dr}{dR} \right|,
 \ee

 \noindent

\noindent  where $f(r) = A(r)/({\rm Volume}[T_{\Delta}])$, and $A(r)$ is given by
 equations (\ref{a1}-\ref{a3}). The distribution $F(R)$ was first determined numerically by
 Zyczkowski et al. in \cite{ZHS98}. Here we compute $F(R)$
 analytically \cite{BCPP02b} and, as expected, the calculations coincide with the 
 concomitant numerical results and the ones reported in \cite{ZHS98}.

\subsection{$\lambda_m$-domain}

Coming back to two-qubits, the quantity $\omega_q$ is not appropriately 
suited to discuss the
limit $q\rightarrow \infty$. However, $\omega_q^{1/q}$ does
exhibit a nice behaviour when $q\rightarrow \infty$. Indeed, we
have

\be \label{limiqinf} \lim_{q\to \infty } \, \left( Tr \rho^q
\right)^{1/q} \, = \,\lim_{q\to \infty } \, \left( \sum_i
p_i^q\right)^{1/q} \, = \, \lambda_m, \ee

\noindent where

\be \lambda_m = \max_{i} \{ p_i \} \ee

\noindent is the maximum eigenvalue of the density matrix $\rho$.
Hence, in the limit $q\to \infty $, the $q$-entropies (when
properly behaving) depend only on the largest eigenvalue of the
density matrix. For example, in the limit $q\to \infty $, the 
R\'{e}nyi entropy reduces to

\be \label{reninf} S^{(R)}_{\infty} \, = \, -\ln \left( \lambda_m
\right).
 \ee

\noindent
 It is worth realizing that the largest eigenvalue itself
constitutes a legitimate measure of mixture. Its extreme values
correspond to (i) pure states ($\lambda_m =1$) and (ii) the
identity matrix ($\lambda_m = 1/4$). It is also interesting to
mention that, for states diagonal in the Bell basis, the
entanglement of formation is completely determined by $\lambda_m$ 
(This is not the case, however, for general states of
two-qubits systems).

 In terms of the geometric representation of the simplex $\Delta$,
 the set of states with a given value $\lambda_m$ of their maximum 
 eigenvalue is represented by the tetrahedron                 
determined by the four planes

 \be \label{geolambda}
  \lambda_m \, = \, 2({\bf r} \cdot {\bf r}_i) \, + \, \frac{1}{4},
  \,\,\,\,\,\, i=1,\ldots, 4.
 \ee

  \noindent
  The four vertices of this tetrahedron are given by the
  intersection points of each one of the four possible triplets
  of planes that can be selected among the four alluded to planes.

For $q \to \infty$ the accessible states with a given degree of
mixture are on the surface of a small tetrahedron $T_l$ concentric
with the tetrahedron $T_{\Delta}$. We are going to characterize
each tetrahedron $T_l$ (representing those states with a given
value of $\lambda_m$) by the distance $l$ between (i) the common
centre of $T_{\Delta} $ and $T_l$ and (ii) each vertex of $T_l$.
The volume associated with states with a value of $\lambda_m$
belonging to a given interval $\lambda_m$ is proportional to the
area $A(l)$ of the portion of $T_l $ lying within $T_{\Delta}$.

Following a similar line of reasoning as the one pursued in the
case $q=2$, we consider three ranges of values for
$l$. The first range of $l$-values is given by $l \in [0, h_1/3]$.
The particular value  $l = h_1/3$ corresponds to a tetrahedron
$T_l$ whose vertices are located at the centres of the faces of
$T_{\Delta}$. Within the  aforementioned range of $l$-values,
$T_l$ is lies completely  within $T_{\Delta}$. Consequently,
$A(l)$ coincides with the area of $T_l$,

\begin{equation}
A_{I}(l) = 24 \sqrt 3 l^2 \label{s12}.
\end{equation}

\noindent The second range of $l$-values corresponds to $l \in
[h_1/3, h_1]$. The area of the part of $T_l$ lying within
$T_{\Delta}$ is now

\begin{equation}
A_{II}(l) =3 \sqrt 3 \left[8l^2 - {3 \over 2} (3l-h_1)^2\right]
\label{s22}
\end{equation}

\noindent Finally, the third range of $l$-values we are going to
consider is $l \in [h_1, 3 h_1]$. In this case we have

\begin{equation}
A_{III}(l) ={3 \over 2} \sqrt 3(3h_1-l)^2 \label{s32}
\end{equation}

\noindent
 In a similar way as in the $q=2$ case, the above
expressions for $A(l)$ lead to the analytical form of the
probability (density) $F(\lambda_m)$ of finding a two-qubits state
with a given value of its greatest eigenvalue,

\be \label{largesteinge}
 F(\lambda_m) \, = \, \frac{A(l)}{{\rm Volume}[T_{\Delta}]} \,
 \left| \frac{dl}{d\lambda_m} \right|.
\ee

\noindent Remarkably enough, as $q$ tends to infinity all discontinuities 
in the derivative of $F(\lambda_m)$ disappear. In 
the $\lambda_m$-domain the distribution is completely smooth, 
as opposed to the $R$-domain.

\section{Analytical approach for 2x3 systems and higher systems}

\subsection{R-domain}

Previously, we obtained the distribution $F(R)$ vs. $R$ for two-qubits 
 ($N=4$), under the assumption that they are distributed according 
 to measure (\ref{memu1}). Through a useful analogy, we have mapped the 
 problem into a geometrical one in ${\cal R}^3$ regarding interior and 
 common sections of two geometrical bodies. In the previous case 
 we saw that the main difficulty lies in the third region, where 
 the region of the growing sphere inside the tetrahedron is not 
 described by a spherical triangle. The extension to higher 
 dimensions, however, requires a thorough account of the geometrical 
 tools required, but still it is in principle possible. 
 So, one can find the distribution of states according to 
 $R = 1/ Tr(\rho^2)$ basically by computing the surface area of a 
 growing ball of radius $r$ in $N-1$ dimensions ({\it sphere}) 
 that remains inside an outer regular $N$-polytope $T_{\Delta}$ 
 ({\it tetrahedron}) of unit length, excluding the common regions. 
 A $(N-1)$-dimensional sphere can be parameterized in cartesian 
 coordinates

\ben \label{vectSphere}
x_1 &=& r \sin(\phi_1) \sin(\phi_2) \sin(\phi_3)\,...\,\sin(\phi_{N-3}) 
\sin(\phi_{N-2}) \nonumber \\
x_2 &=& r \sin(\phi_1) \sin(\phi_2) \sin(\phi_3)\,...\,\sin(\phi_{N-3}) 
\cos(\phi_{N-2}) \nonumber \\
x_3 &=& r \sin(\phi_1) \sin(\phi_2) \sin(\phi_3)\,...\,\cos(\phi_{N-3}) 
\nonumber \\
...\nonumber \\
x_{N-2} &=& r \sin(\phi_1) \cos(\phi_2) \nonumber \\
x_{N-1} &=& r \cos(\phi_1),
\een

\noindent with the domains $0 \le \phi_j \le \pi$ for $1 \le j \le N-3$ and 
$0 \le \phi_{N-2} < 2\pi$. The definition of the 
$N$-polytope $T_{\Delta}$ then is required. This problem is not 
trivial at all, because new geometrical situations appear in the intersection 
of these two bodies. In point of fact there are $N-2$ intermediate 
regimes between $R=N$ and $R=1$ appearing at integer 
values of $R$ (recall the previous two-qubits case), where a change 
in the growth of interior hyper-surfaces occurs (at the values 
$r_i=\sqrt{(N-R_i)/2R_iN}, R_i=1,...,N$). In any case we can 
always generate random states $\rho$ in arbitrary dimensions 
and compute the corresponding $F_N(R)$ distributions. This is done 
in Fig. 2 for several cases.                    	                     
The relation (\ref{radiopart}) is generalized to $N$ dimensions in the form

\begin{figure}[ht]
\centering   \includegraphics[scale= 0.6]{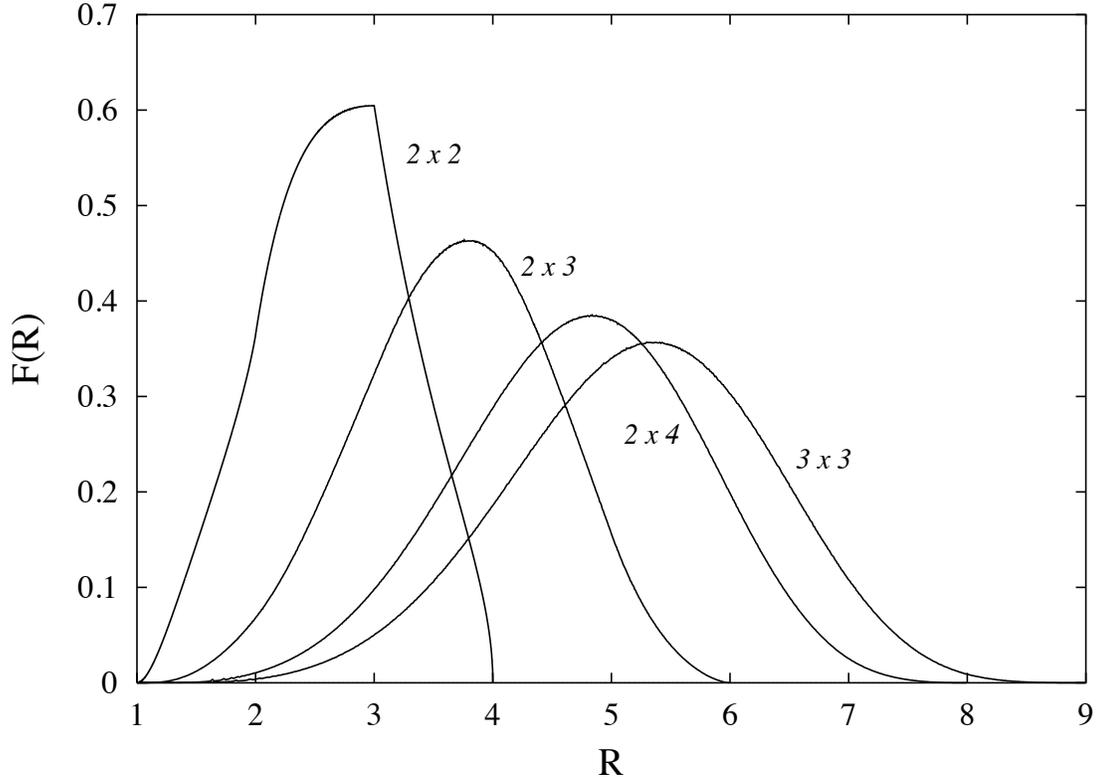}
\caption{Plot of the $F_N(R)$ distributions of mixed states $\rho$ numerically 
computed in arbitrary dimensions, generated according to (\ref{memu}). 
As we increase the total dimension $N$, the curves become smoother, 
in correspondence with our geometric interpretation. See text for details.}
\end{figure}

\be \label{rRN}
r^2 = -  {1 \over 2N} + {1 \over 2 R}.
\ee

\noindent The distribution $F_N(R), \, R \in [N-1,N]$ can be obtained 
analytically 

\be
F_N(R) \, \sim \, \frac{1}{R^2}\big[ \frac{1}{R}-\frac{1}{N} 
\big]^{\frac{N-3}{2}}
\ee

\noindent which has been numerically checked. 
The particular form of $F(R)$ for arbitrary $N$ is difficult 
to obtain, but nevertheless one can obtain quantitative results for 
asymptotic values of $N$. It may be interesting to know the position 
of the maximum of $F(R)$ or the mean value $\langle R \rangle$, which turns 
to be $\simeq N/2$ \cite{Zyckasimp} for states $\rho$ generated 
according to (\ref{memu1}). There is the so called Borel lemma \cite{Borel} 
in discrete mathematics that asserts that (translated to our problem) 
when you grow a $(N-1)$-ball inside $T_{\Delta}$, from the moment that it 
swallows, say, 1/2 of the volume of it, then the area outside drops very 
quickly with further grow. So the maximum intersection with the sphere should 
be approximately for the radious $r^{*}$ where the volume of the ball $V_{N-1}$ 
{\it equals} that of the $T_{\Delta}$-polytope $V_T$. 
The usual formulas for the volumes of ($N-1$)-dimensional spheres and regular unit $N$-polytopes are 
$V_{N-1}=\frac{\pi^{(N-1)/2}}{\Gamma(\frac{N-1}{2}+1)}r^{N-1}$ and 
$V_T=\frac{1}{(N-1)!}\sqrt{\frac{N}{2^{N-1}}}$, respectively. 

It is then that we can assume 
that the position of $R(r^{*}) \simeq R^{\prime}$ such that $F(R=R^{\prime})$ 
is maximal. Substituting $r^{*}$ in (\ref{rRN}), and after some algebra, 
we obtain the beautiful result

\be \label{deltaR}
\lim_{N \to \infty} \, \frac{1/R(r^{*})}{1/N} \, = \, \frac{2\pi+e}{2\pi} 
\, \simeq \, 1.43.
\ee

\noindent In other words, $F_N(1/R) \sim \delta(1/R-1/N)$ for large $N$.

We must emphasize that this type of distributions $F_N(R)$ are 
``degenerated" in some cases, that is, different systems may 
present identical $F(R)$ distributions (for instance, there is nothing 
different from this perspective between $2\times 6$ and 
$3\times 4$ systems). We do not know to what extend these 
distributions are physically representative of such cases, as far 
as entanglement is concerned. What is certain is that all 
states $\rho$ with $R \in [N-1,N]$ possess a positive partial transpose. In 
point of fact, they are indeed separable, as shown in \cite{balls}.  
We merely mean by this that a state close enough to the maximally mixed state 
$\frac{1}{N}I_N$ is always separable. In other words, states lying on 
$(N-1)$-spheres with radius $r \le r_c\equiv 1/\sqrt{2N(N-1)}$ 
are always separable.

\subsection{$\lambda_m$-domain}

When regarding the maximum 
 eigenvalue $\lambda_m$ as a proper degree of mixture one is 
 able to find a geometrical picture analogue to the one of the 
 growing sphere. In that case a nested inverted tetrahedron 
 grows inside the outer tetrahedron representing the simplex 
 of eigenvalues $\Delta$. The generalization to higher bipartite 
 systems is similar to the $R$-case, but far much easier to 
 implement mathematically. As in that case, we have a high degree 
 of symmetry in the problem. The advantage is that one does not 
 deal with curved figures but perfectly flat and sharp surfaces instead. This 
 fact makes the general problem more approachable.

 We have seen that the problem of finding how the states of a bipartite quantum 
mechanical system are distributed according to their degree of mixedness 
can be translated to the realm of discrete mathematics. 
If we consider our measure of mixedness to be the maximum eigenvalue $\lambda_m$ 
of the density matrix $\hat \rho$ and the dimension of our problem to be 
$N=N_A \times N_B$, we compute the distribution of states in arbitrary dimensions 
by letting an inner regular $N$-polytope $T_l$ to grow inside an outer unit length 
$N$-polytope $T_{\Delta}$, the vertices of the former pointing towards the 
centre of the faces of the latter. In fact, it can be shown that the radius 
$l$ of the maximum hypersphere that can be inscribed inside the inner 
polytope is directly related to $\lambda_m$. 

By computing the surface area of $T_l$ 
strictly inside $T_{\Delta}$, we basically find the desired probability 
(density) $F_N(\lambda_m)$ of finding a state $\hat \rho$ with maximum 
eigenvalue $\lambda_m$ in $N$ dimensions. 

To fix ideas, it will prove useful first to define the vertices of $T_{\Delta}$ 
and $T_l$. In fact it is essential, because we need to deal with elements of 
cartesian geometry in $N$-dimensions. This vectors are given as

\begin{eqnarray}
\vec{r_1} &=& (-\frac{1}{2},-\frac{1}{2\sqrt{3}},-\frac{1}{4}\sqrt{\frac{2}{3}},...,-\frac{1}{N-1}\sqrt{\frac{N-1}{2N}}) \nonumber \\
\vec{r_2} &=& (\frac{1}{2},-\frac{1}{2\sqrt{3}},-\frac{1}{4}\sqrt{\frac{2}{3}},...,-\frac{1}{N-1}\sqrt{\frac{N-1}{2N}}) \nonumber \\
\vec{r_3} &=& (0,\frac{1}{\sqrt{3}},-\frac{1}{4}\sqrt{\frac{2}{3}},...,-\frac{1}{N-1}\sqrt{\frac{N-1}{2N}}) \nonumber \\
\vec{r_4} &=& (0,0,\frac{3}{4}\sqrt{\frac{2}{3}},..,-\frac{1}{N-1}\sqrt{\frac{N-1}{2N}}) \nonumber \\
...& & \nonumber \\
\vec{r_N} &=& (0,0,0,...,\sqrt{\frac{N-1}{2N}}),\label{vectr}
\end{eqnarray}

\noindent with $\sqrt{\frac{N-1}{2N}}$ being the distance from the center 
to any vertex of this regular $N$-polytope of unit length. One can easily check 
that $\sum_{i} \vec{r_i}=\sum_{i,j} \vec{r_i} \cdot \vec{r_j}=0$, as required. 
This particular choice for the position of the vertices of this $N$-simplex is 
such that it simplifies going from one dimension to the next by adding a new 
azimuthal axis each time. This vectors comply with the relations

\ben
\vec{r_i}\cdot\vec{r_j}\, &=& \, -\frac{1}{2N}+\frac{1}{2}\delta_{ij}, \cr
\lambda_m\, &=& \, 2(\vec{r}\cdot\vec{r_i}) + \frac{1}{N}, \,\,\,\,\,\,\,\, 
i=1...N,
\een 

\noindent where the last equation is the general form of (\ref{geolambda}).

Once we have a well defined $T_{\Delta}$, to know the coordinates of $T_l$ is 
straightforward. In fact, $T_l$ is the reciprocication (see \cite{Sommer}) 
of $T_{\Delta}$. This means that the coordinates of $T_l$ are obtained by 
reversing the sign of the ones of $T_{\Delta}$, multiplied by a suitable 
factor (which can be shown to be $\sqrt{2N(N-1)}l$, with $l$ defined as the 
length between the centre of $T_l$ to the centre of any of its faces, which 
in turn points towards the vertices of $T_{\Delta}$). Thus, we can 
relate $l$ with $\lambda_m$ through a general (\ref{geolambda})-relation 
$\lambda_m=2\,l\,\sqrt{\frac{N-1}{2N}}+\frac{1}{N}$, 
such that $\frac{dl}{d\lambda_m}=\sqrt{(2N)/(N-1)}/2$.

\begin{figure}[ht]
\centering   \includegraphics[scale= 0.6]{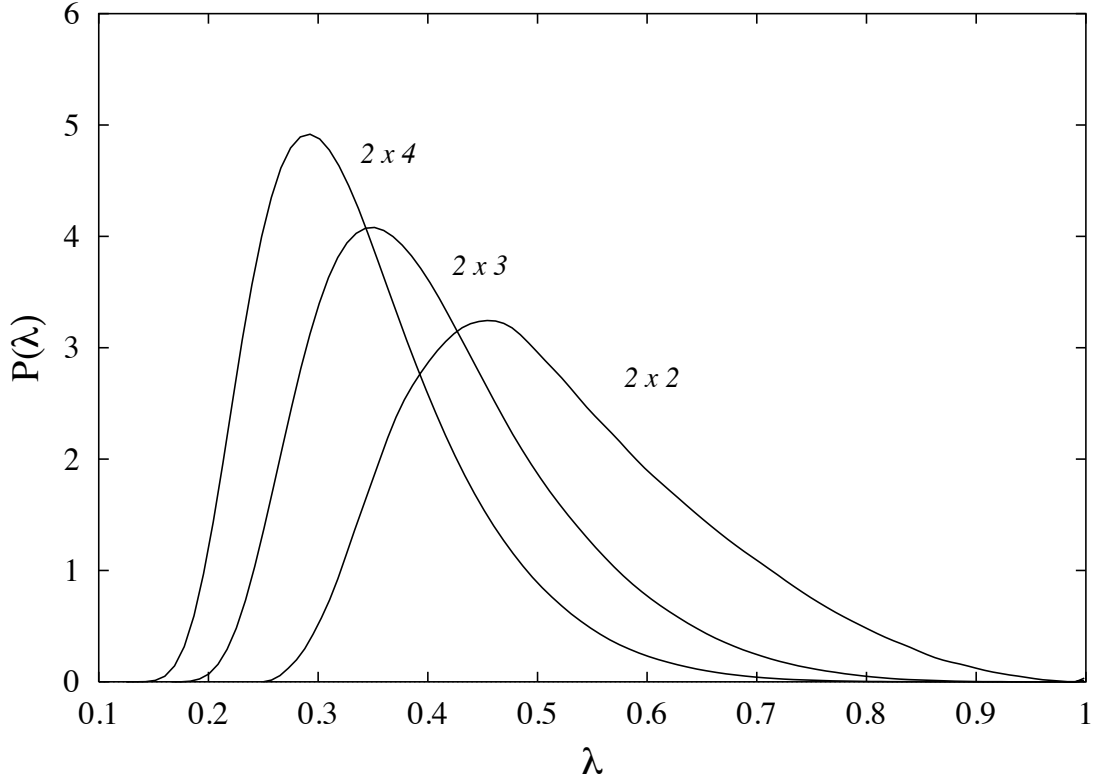}
\caption{Plot of the $F_N(\lambda_m)$ distributions of mixed states $\rho$ numerically 
computed in arbitrary dimensions, generated according to (\ref{memu1}). 
As we increase the total dimension $N$, the curves tend to peak around 
$1/N$. See text for details.}
\end{figure}

Several distributions $F_N(\lambda_m)$ are obtained numerically by generating 
random states $\rho$ according to (\ref{memu1}) in Fig. 3. 
It becomes apparent that as $N$ grows, the distributions 
are biased towards $\lambda \simeq 1/N$, in absolute agreement with 
the result (\ref{deltaR}). 

As in the $R$-case, $F_N(\lambda_m)$ is distributed into $N-1$ regions 
separated at fixed values of $\lambda_m^{(i)}=\frac{1}{N-i}, \, i=1..(N-2)$. 
The general recipe for obtaining $F_N(\lambda_m)$ is tedious and long, but 
some nice general results are obtained. The $F_N(\lambda_m)$-distributions 
for the ranges a) $\lambda_m \in [\frac{1}{N},\frac{1}{N-1}]$ and 
b) $\lambda_m \in [\frac{1}{2},1]$ are general and read

\ben \label{region}
F_I(\lambda_m) \, &=& \, \kappa \frac{N}{(N-2)!}
\sqrt{\frac{N-1}{2^{N-2}}} \big[\sqrt{2N(N-1)}l(\lambda_m) \big]^{N-2}, \cr
F_{Last}(\lambda_m) \, &=& \, \kappa \frac{N}{(N-2)!}
\sqrt{\frac{N-1}{2^{N-2}}} \bigg[ \frac{\sqrt{\frac{N-1}{2N}}-l(\lambda_m)}
{\sqrt{ \frac{N}{2(N-1)} }} \bigg]^{N-2},
\een

\noindent respectively, 
where $\kappa\equiv\frac{dl}{d\lambda_m}/{\rm Volume}[T_{\Delta}]$ 
is introduced for convenience. 

For the qubit-qutrit system, we have $(N=6)$. Defining $r_i \equiv \sqrt{\frac{N-1}{2N}}$ 
and $y_i \equiv (l(\lambda_m)(N-1)-\frac{i}{N-i}r_i)/(\sqrt{N/2(N-1)})$, 
in addition to the previous regions (\ref{region}) we obtain 

\ben
F_{II}(\lambda_m) \, &=& \, \kappa \big[F_I(\lambda_m)/\kappa -
\frac{(N-1)N}{(N-2)!}\sqrt{\frac{N-1}{2^{N-2}}} [y_{i=1}]^{N-2} \big]; \cr
F_{III}(\lambda_m) \, &=& \, \kappa \big[F_{II}(\lambda_m)/\kappa+ 
\frac{2^9}{5^4}\frac{(N-1)N}{(N-2)!}\sqrt{\frac{N-1}{2^{N-2}}} [y_{i=2}]^{N-2} \big]; \cr
F_{IV}(\lambda_m) \, &=& \, \kappa \big[F_{III}(\lambda_m)/\kappa- 
2\frac{3^4}{5^4}\frac{(N-1)N}{(N-2)!}\sqrt{\frac{N-1}{2^{N-2}}} [y_{i=3}]^{N-2} \big],
\een

\noindent for $\lambda_m \in [\frac{1}{5},\frac{1}{4}],[\frac{1}{4},\frac{1}{3}]$, and 
$[\frac{1}{3},\frac{1}{2}]$, respectively. From the previous formulas one can 
infer a general induction procedure. 
Analytical results are in excellent agreement with numerical generations.

\section{Non-uniqueness of a general measure for mixed states. Geometric implications}

In Refs. \cite{ZHS98,Z99}, 
a basic question regarding a natural measure $\mu$ for the set of mixed states 
$\rho$ was debated. As described in Secs. (7.1) and (9.1), it is know, the set 
of all states ${\cal S}$ can 
be regarded as the cartesian product ${\cal S} = {\cal P} \times \Delta$, 
where $\cal P$ stands for the family of all
complete sets of ortonormal projectors $\{ \hat P_i\}_{i=1}^N$,
$\sum_i \hat P_i = I$ ($I$ being the identity matrix), and $\Delta$
is the set of all real $N$-tuples $\{\lambda_1, \ldots, \lambda_N
\}$, with $\lambda_i \ge 1$ and $\sum_i \lambda_i = 1$. It is universally accepted to assume            
the Haar measure $\nu$ to be the one defined over ${\cal P}$, because of 
its rotationally-invariant properties. But when it turns to discuss an 
appropriate measure over the simplex $\Delta$, some controversy arises. 
In all previous considerations here, we have regarded the Lebesgue measure 
${\cal L}_{N-1}$ as being the ``natural" one. But one must mention that 
Slater has argued \cite{sla1,sla2} that, in analogy
to the classical use of the volume element of the Fisher information
metric as Jeffreys' prior \cite{jef} in Bayesian theory, a natural measure
on the quantum states would be the volume element of the Bures
metric. The problem lies on the fact that there is no unique 
probability distribution defined over the simplex of eigenvalues $\Delta$ of 
mixed states $\rho$. In point of fact, the debate was motivated by the fact 
that the volume occupied by separable two-qubits states was     
found in \cite{ZHS98} to be greater than $50\%$ ($P_{sep}=0.6312$) using 
the measure $\mu$, something which is surprising.

One such probability distribution that is suitable for general 
considerations is the Dirichlet distribution \cite{Z99}

\be \label{dirich}
P_{\eta}(\lambda_1, \ldots, \lambda_N) \,=\, 
C_{\eta} \lambda_1^{\eta-1}\lambda_2^{\eta-1}...\lambda_N^{\eta-1},
\ee

\noindent with $\eta$ being a real parameter and 
$C_{\eta}=\frac{\Gamma[N\eta]}{\Gamma[\eta]^N}$ the normalization 
constant. This is a particular case of the more general Dirichlet distribution. 
The concomitant probability density for variables $(\lambda_1,...,\lambda_N)$ 
with parameters $(\eta_1,...,\eta_N)$ is defined by 

\be \label{dirich2}
P_{{\bf \eta}}(\lambda_1, \ldots, \lambda_N) \,=\, 
C_{\bf {\eta}} \lambda_1^{\eta_1-1}\lambda_2^{\eta_2-1}...\lambda_N^{\eta_N-1},
\ee 

\noindent with $\lambda_i \ge 0$, $\sum_{i=1}^{N} \lambda_i=1$ and 
$\eta_1,...,\eta_N > 0$, and $C_{\bf {\eta}}=\Gamma(\sum_{i=1}^N \eta_i)/
\prod_{i=1}^N \Gamma(\eta_i)$. Clearly, distribution (\ref{dirich2}) generalizes 
(\ref{dirich}). This distribution admits 
a clear interpretation. As known, the multinomial distribution provides a probability of 
choosing a given collection of $M$ items out of a set of $N$ items with repetitions, 
the probabilities being $(\lambda_1,...,\lambda_N)$. These probabilities are the parameters of 
the multinomial distribution. The Dirichlet distribution is the conjugate prior of the parameters 
of the multinomial distribution.

A new measure then can be defined as 
$\mu_{\eta}=\nu\times \Delta_{\eta}$, where $\Delta_{\eta}$ denotes the simplex 
of eigenvalues distributed according to ({\ref{dirich}}) 
(The Haar measure $\nu$ remains untouched). Thus, one clearly recovers 
the Lebesgue measure ${\cal L}_{N-1}$ for $\eta=1$ (uniform distribution), 
and Slater's argumentation reduces to 
take $\eta=\frac{1}{2}$ in ({\ref{dirich}}). For $\eta \rightarrow 0$ one 
obtains a singular distribution
concentrated on the pure states only, while for $\eta \rightarrow \infty$, 
the distribution peaks on the maximally mixed state $\frac{1}{N}I$. We will 
see shortly that changing the continuous parameter $\eta$ indeed modificates 
the average purity (as expressed in terms of $R=1/Tr(\rho^2)$) 
of the generated mixed states. 

In what follows we numerically generate mixed states whose eigenvalues are 
distributed following ({\ref{dirich}}). This is done is order to tackle 
the dependence of relevant quantities on the parameter $\eta$. Let us 
consider the way mixed states are distributed according to $R$. We focus our 
attention on the two-qubits instance, but similar studies can be extended 
to arbitrary bipartite dimensions. 
As shown in Fig. 4, the distributions $P(R)$ vs. $R$ are shown for 
$\eta=\frac{1}{2},1,2$ (from left to right in this order) while Fig. 5 shows 
analogous distributions for the maximum eigenvalue 
$\lambda_m$ for $\eta=\frac{1}{2},1,2$ (from right to left). Notice the different shapes. 
We can no longer attribute a 
geometrical description to $P(R)$ except for $\eta=1$. 
In \cite{Z99} $P(R)$ for $\eta=\frac{1}{2}$ was first derived. Here we can 
provide different distributions for arbitrary $\eta$-values.

\begin{figure}[ht]
\centering   \includegraphics[scale= 0.6]{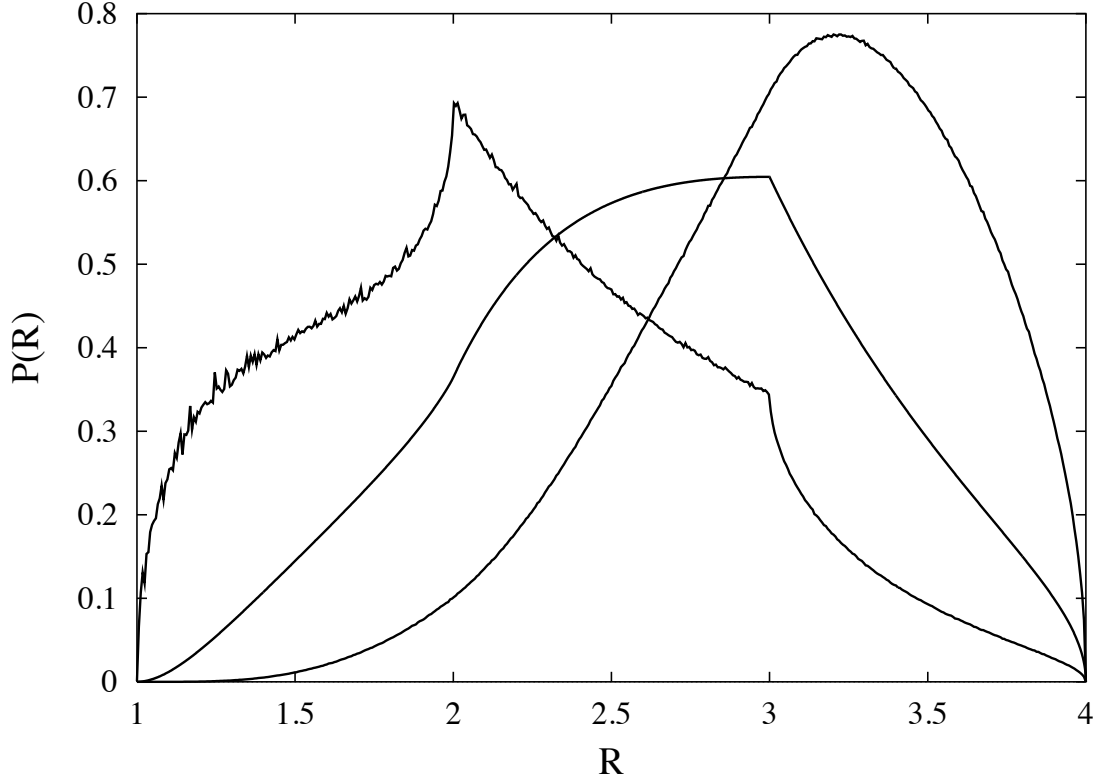}
\caption{$P(R)$ vs. $R$ distributions for two qubit systems, 
whose eigenvalues are distributed according to ({\ref{dirich}}), for the values 
$\eta=\frac{1}{2},1,2$ (from left to right in this order). It is plain 
from this figure that the uniform distribution ($\eta=1$) appears 
more balanced that the others. Also, the particularity of $R=2,3$ seems to disappear 
for $\eta > 1$. See text for details.}
\end{figure}

\begin{figure}[ht]
\centering   \includegraphics[scale= 0.6]{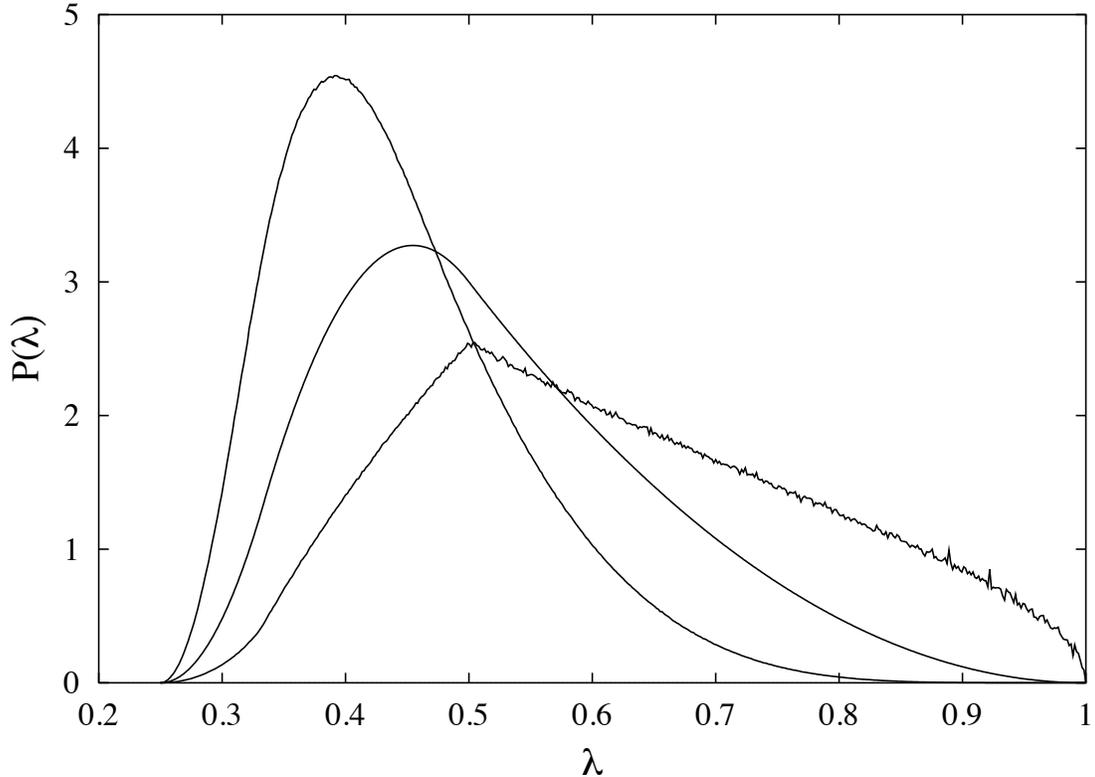}
\caption{Probability (density) distributions of the maximum eigenvalue $\lambda_m$ 
of two qubit systems, whose eigenvalues are distributed according to 
({\ref{dirich}}), for the values 
$\eta=\frac{1}{2},1,2$ (from right to left). When employing $\lambda_m$ as a 
degree of mixture, the derivative of these distributions is discontinuous at the 
special values $\lambda_m=\frac{1}{2},\frac{1}{3}$ for $\eta < 1$. See text for details.}
\end{figure}

A way to devise a certain range of reasonable $\eta$-values is to study the 
average $R$ induced for every $\eta$-distribution. This is performed in 
Fig. 6. The average $R$-value $\langle 1/Tr(\rho^2) \rangle$ and 
$R^{*} \equiv 1/\langle Tr(\rho^2) \rangle$ are plotted versus $\eta$. 
$\langle R \rangle$ (solid line) can only be computed numerically, 
but luckily $R^{*}$ (dashed line) is obtained in analytical fashion 
{\it for all N}

\ben \label{traceN}
\langle {\rm Tr} \rho^2 \rangle_N(\eta) &=& C_{\eta} 
\int_{0}^{1}d\lambda_1\lambda_1^{\eta-1}
\int_{0}^{1-\lambda_1}d\lambda_2\lambda_{2}^{\eta-1}...
\int_{0}^{1-\sum_{i=1}^{N-2}\lambda_i}d\lambda_{N-1}\lambda_{N-1}^{\eta-1} \cr 
& &(1-\sum_{i=1}^{N-1}\lambda_i)^{\eta-1}\,\big[\sum_{j=1}^{N} 
\lambda_{j}^{2}\big]\, = \, \bigg[N - \frac{N-1}{\eta+1} \bigg]^{-1}.
\een

\noindent The fact that $R^{*}$ matches exact results 
validates all our present generations. The actual value 
$\langle R \rangle$ is slightly larger than $R^{*}$ for all values 
of $\eta$, but both of them coincide for low and high values of the 
parameter $\eta$. It is obvious from Fig. 6 that we cannot 
choose distributions that depart considerably from the uniform one $\eta=1$, 
because in that case we induce probability distributions that favor high or 
low $R$ already.

\begin{figure}[ht]
\centering   \includegraphics[scale= 0.6]{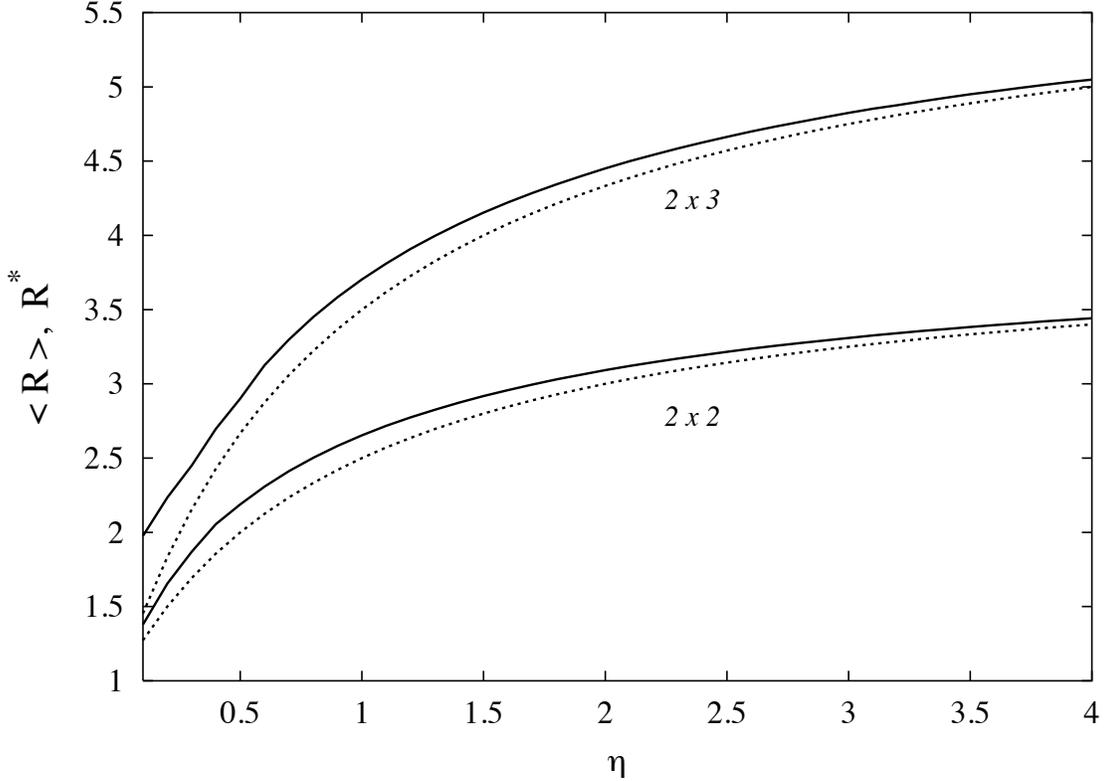}
\caption{Average $R$-value $\langle 1/Tr(\rho^2) \rangle$ (solid line) and 
$R^{*} \equiv 1/\langle Tr(\rho^2) \rangle$ (dashed line) for 
two qubit and one qubit-qutrit systems, plotted versus the Dirichlet 
parameter $\eta$. See text for details.}
\end{figure}

\begin{figure}[ht]
\centering   \includegraphics[scale= 0.6]{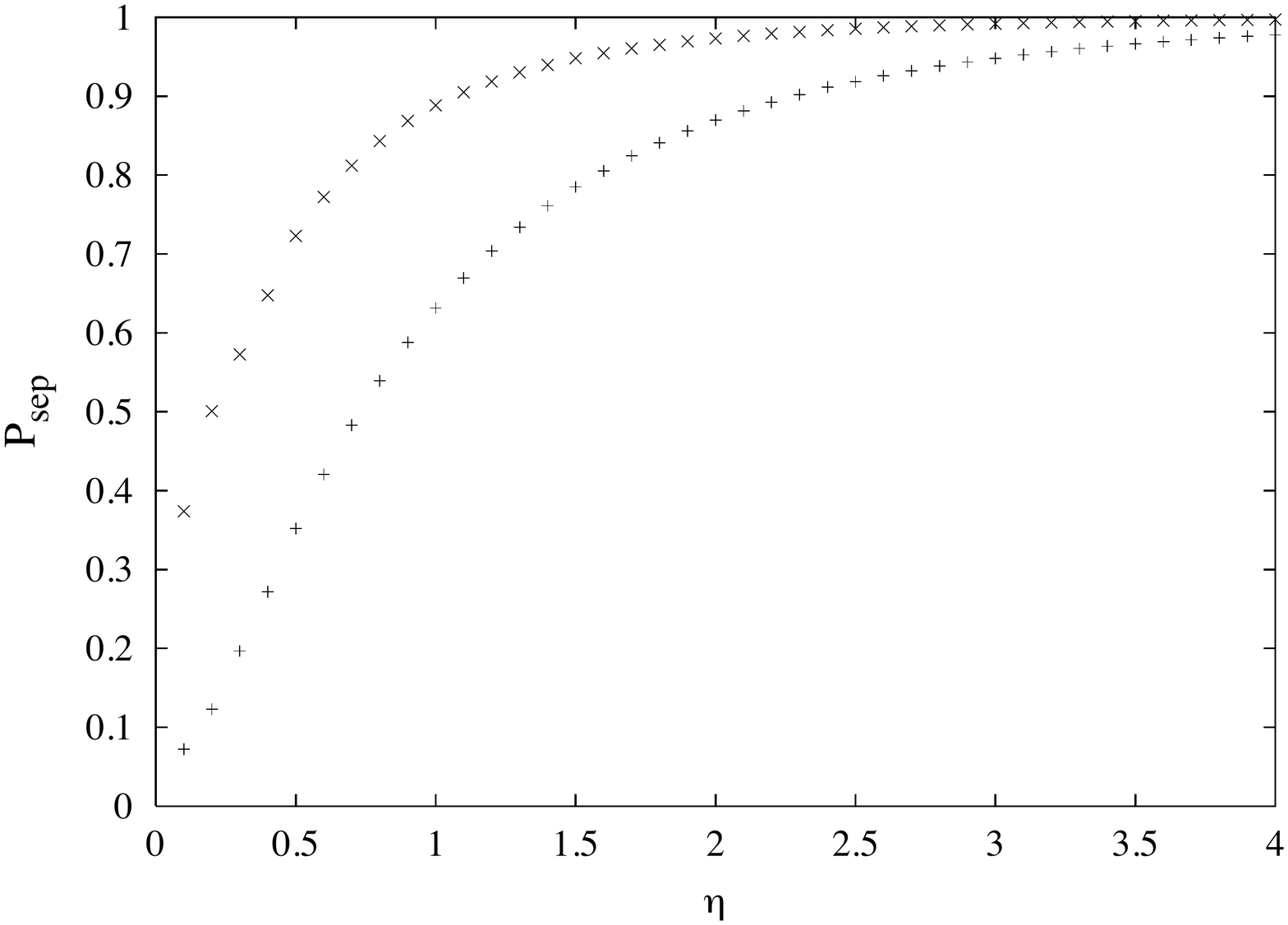}
\caption{Probability of finding a state $\rho$ of two-qubits being positive 
partial transposed (lower curve), and violating the strongest entropic 
criterion $q=\infty$ (upper curve). This figure illustrates the fact that 
one can arbitrarily choose any $P_{sep}$ by generating two qubit mixed states 
with different Dirichlet parameter $\eta$. See text for details.}
\end{figure}

Perhaps the best way is to go straight to the question that originated 
the controversy on the $\Delta$-measures: what is the dependency of 
the {\it a priori} probability $P_{sep}$ of finding a two-qubits mixed state 
being separable? In Fig. 7 we depict $P_{sep}$ vs. $\eta$ for 
states complying with PPT (lower curve) and those which violate the $q=\infty$-entropic 
inequalities (upper curve). It seems reasonable to assume that a permissible range 
of $\eta$-distributions belong to the interval $[\frac{1}{2},2]$, within 
which $P_{sep}$ remains around the reference point $P_{sep}=0.5$. 

However, in view of the previous outcomes we believe that the 
results obtained considering the uniform 
$\eta=1$-distribution for the simplex $\Delta$ (the one that allow us to exploit a simple geometric analogy) 
remains the most natural 
choice possible, independent of any form that one may adopt for a generic 
probability distribution.

\section{Conclusions}

We have introduced a geometric picture to obtain the probability $F$ of finding quantum states of 
two-qubits with a given degree of mixture (as measured by an 
appropriate function of $\omega_q$) is analytically found for $q=2$ 
and $q \to\infty$. In the latter case, the $q$-entropies become functions
of the statistical operator's largest eigenvalue $\lambda_m$. In
point of fact, $\lambda_m$ itself constitutes a legitimate measure
of mixture. During the derivation of the probability (density) distributions 
$F_N$ of finding a bipartite mixed state in arbitrary dimensions 
$N=N_A\times N_B$ with a given degree of mixture, we saw that it is more 
convenient to use $\lambda_m$ instead of $R$. 
Finally, we have derived explicitly the 
distribution $F_N(\lambda_m)$ vs. $\lambda_m$ for the physical meaningful 
case of a qubit-qutrit system ($N=6$). 

\section*{Acknowledgements}

J. Batle acknowledges partial support from the Physics Department, UIB. 
J. Batle acknowledges fruitful discussions with J. Rossell\'o, Maria del Mar Batle and Regina Batle.

\end{document}